\documentclass[twocolumn,showpacs,preprintnumbers]{revtex4}
\usepackage{amsmath}
\usepackage{dcolumn}
\usepackage{bm}
\usepackage{graphicx}
\usepackage{amsfonts}
\usepackage{amssymb}
\usepackage{mathrsfs}
\usepackage{color}

\begin{document}

\title{The vacuum induced Berry phase beyond rotating-wave approximation}
\author{Tao Liu$^{1,2}$, Mang Feng$^{2}$ \footnote{E-mail: mangfeng@wipm.ac.cn} and Kelin Wang$^{3}$}
\affiliation{$^{1}$ The School of Science, Southwest University
of Science and Technology, Mianyang 621010, China \\
$^{2}$ State Key Laboratory of Magnetic Resonance and
Atomic and Molecular Physics and Key Laboratory of Atomic Frequency Standards, Wuhan Institute of Physics and
Mathematics, Chinese Academy of Sciences, and Wuhan National
Laboratory for Optoelectronics, Wuhan, 430071, China \\
$^{3}$ The Department of Modern Physics, University of Science and
Technology of China, Hefei 230026, China}

\begin{abstract}
With reference to the vacuum induced Berry phase (VIBP) obtained in the interaction
of a spin-1/2 particle with quantized irradiation field under rotating-wave approximation
(RWA), we present completely different treatment for the VIBP by a fully quantum
mechanical treatment beyond the RWA, which gives a new definition of the VIBP and indicates
the validity of RWA to be relevant
not only to the energy conservation and comparison of characteristic parameters, but
also to more subtle physics, such as the geometric property of the state evolution.
Our result is of conceptual importance and also of significant relevance to quantum information processing.
\end{abstract}

\maketitle

The Berry phase presents the geometric and topological property of
quantum system under an adiabatic cyclic evolution \cite {berry}, which
has been widely investigated over past decades from the viewpoints
of fundamental physics and the application for quantum information
processing (QIP). A recent work \cite {vacuum} on Berry phase by a fully quantum
mechanical treatment demonstrated the existence of the Berry phase
relevant to vacuum field, which could be tested by a single spin-1/2 interacting
with a quantized irradiation field in terms of Jaynes-Cummings (JC) model \cite
{jc}. This vacuum induced Berry phase (VIBP), due to no counterpart
in classical treatment, has brought about a series of studies for
more interesting physical interpretation and application.

We argue in this Letter that the VIBP defined in \cite {vacuum} strongly depends
on the employed rotating-wave approximation (RWA). Despite success
for tens of years for light-matter interaction, the
JC model has been debated due to employment of the RWA \cite {non-RWA},
although the RWA has been justified from the viewpoint of energy
conservation and works very well if the light-matter coupling is
weak enough compared to other characteristic parameters in the
system. Our remarkable results include: (1) The VIBP defined in
\cite {vacuum} is actually relevant to both the vacuum state and the one-boson state
of the quantized field. We thereby define a new VIBP which is only associated with vacuum quantized field.
We show this newly defined VIBP to be constantly zero under the RWA but different from zero or
any integer multiple of 2$\pi$ beyond the RWA; (2)
With reference to the VIBP defined in \cite {vacuum}, the
counterparts beyond the RWA are actually the Berry phases for the lowest
excited states, whose values are much different from those in \cite
{vacuum}. This reminds us to pay more attention to identification of
the generated Berry phase, particularly in some recent schemes for
QIP using geometric phases based on JC model \cite {qip1}.

The key point of our work is to solve the JC model in the absence of the
RWA by a method called coherent-state diagonalization \cite {TLiu}, which enables
us to obtain eigensolutions in a very good approximation and to
compare with the results under the RWA one by one. In contrast to
the conventional viewpoint that validity of the RWA depends on the
energy conservation and the strength comparison of the parameters in
the system, we demonstrate that breakdown of the RWA is related to  more
subtle physics, such as the geometric property of the state
evolution.

We first review briefly the idea in Ref. \cite{vacuum} by
considering a two-level system interacting with a single-mode
quantum field under the RWA, written in units of $\hbar$ as,
\begin{equation}
H_{R}=\frac{\omega_{0}}{2}\sigma_{z} + \omega a^{\dagger}a +
g(a^{\dagger}\sigma_{-} + a\sigma_{+}),
\end{equation}
where $\omega_{0}$ is the resonant frequency of the two levels (i.e., the upper level $|e\rangle$
and the lower level $|g\rangle$),
and $\omega$ and $a$ ($a^{\dagger}$) are, respectively, the frequency and the
annihilation (creation) operator of the single mode. $g$ means the
coupling and $\sigma_{z,+,-}$ are usual Pauli operators. Under the
rotation with respect to $\omega(a^{\dagger}a + \sigma_{z}/2)$ and then under
a unitary transformation $U(\varphi)=e^{-i\varphi
a^{\dagger}a}$, Eq. (1) is rewritten as,
\begin{equation}
H'_{R}=\frac{\Delta}{2}\sigma_{z} + g(a^{\dagger}e^{-i\varphi}\sigma_{-}
+ ae^{i\varphi}\sigma_{+}),
\end{equation}
with $\Delta=\omega_{0}-\omega$. Considering the eigenstates
$|\Psi_{n}^{\pm}(\varphi)\rangle$ of Eq. (2) to traverse a loop $C$
on the associated Bloch sphere with $\varphi$ varied adiabatically from 0
to $2\pi$, we obtain the Berry phase $\gamma_{n,\pm}=i\int_{C}
d\varphi \langle\Psi_{n,\pm}(\varphi)|d/d\varphi
|\Psi_{n,\pm}(\varphi)\rangle$ to be
\begin{equation}
\gamma_{n,\pm}=\pi(1\mp\cos\theta_{n})+2\pi n,    \notag
\end{equation}
with $\cos\theta_{n}=\Delta/\sqrt{\Delta^{2}+4g^{2}(n+1)}$. This implies
the VIBP defined in \cite {vacuum} to be different from zero or
any integer multiple of 2$\pi$ in the case of $n=0$.
In above treatment, we have used the eigensolutions $|\Psi_{n,\pm}(\varphi)\rangle
=e^{i\varphi a^{+}a}|\Psi_{n,\pm}\rangle$ with
\begin{equation}
|\Psi_{n,\pm}\rangle = c_{n,\pm}|n\rangle|e\rangle + d_{n,\pm}|n+1\rangle|g\rangle,
\end{equation}
where $c_{n,+}=\cos\theta_{n}/2$, $d_{n,+}=\sin\theta_{n}/2$,
$c_{n,-}=\sin\theta_{n}/2$ and $d_{n,-}=-\cos\theta_{n}/2$. As
shown in Appendix, the Hamiltonian in Eq. (1) commutes with the parity operator
$P=-\sigma_{z} e^{i\pi a^{\dagger}a}$, and thereby $\gamma_{0,\pm}$, the VIBPs defined
in \cite {vacuum}, both correspond to the wavefunctions of odd parity. This will
be more clarified in our later comparison with the
treatment beyond the RWA. Another point we have to emphasize is the
eigenfunction $|\Psi_{GS}\rangle=|0\rangle|g\rangle$ omitted in Eq.
(3), which was conventionally considered as a trivial solution due to
decoupling from the interaction. In fact, $|\Psi_{GS}\rangle$ is energetically lower
than $|\Psi_{0,\pm}\rangle$ and should thereby be the real ground
state of the system. More importantly, it is only associated with the vacuum quantized field.
Instead, $|\Psi_{0,\pm}\rangle$ employed in \cite {vacuum} to define the VIBP is actually related to both
$|0\rangle$ and $|1\rangle$ of the field. In this sense, we prefer to define the VIBP in the present Letter based on
$|\Psi_{GS}\rangle$, and call the VIBP defined in \cite {vacuum} as original VIBP. It is easy to obtain our defined
VIBP $\gamma_{GS}=0$ which implies no VIBP under RWA.
This consideration would be more interesting when we go beyond the RWA below.

Elimination of the RWA from Eq. (1) yields
\begin{equation}
H_{NR}=\frac{\omega_{0}}{2}\sigma_{z} + \omega a^{\dagger}a +
g(a^{\dagger}+ a)\sigma_{x},
\end{equation}
where the appearance of the counter-rotating terms
$\sigma_{+}a^{\dagger}$ and  $\sigma_{-}a$ makes it harder to solve
Eq. (4) exactly. Conventional approaches to Eq. (4) have been
developed under different approximations yielding solutions within
some particular ranges of parameters. What we adopt in this Letter
is the solution by diagonalizing Eq. (4) using displaced coherent states \cite
{TLiu}, which could obtain eigensolutions of Eq. (4) in a very good
approximation.  For convenience to employ the results from \cite
{TLiu}, we first perform a unitary transformation on Eq. (4) by
$V=e^{-i\pi\sigma_{y}/4}$, which yields
\begin{equation}
H_{NRX}=-\frac{\omega_{0}}{2}\sigma_{x} + \omega a^{\dagger}a +
g(a^{\dagger}+ a)\sigma_{z}.
\end{equation}
Eq. (5) has the same eigenenergies as Eq. (4) and owns the
eigenfunctions  written as \cite {TLiu},
\begin{equation}
|\Phi^{\pm}\rangle = \frac {1}{\sqrt{2}} \sum_{n}f_{n}[|n\rangle_{\alpha}|e\rangle \pm (-1)^{n}|n\rangle_{-\alpha}|g\rangle],
\end{equation}
with $\alpha=g/\omega$ and the coefficient $f_{n}$ to be determined.
$|n\rangle_{\alpha}$ ($|n\rangle_{-\alpha}$) is a kind of displaced coherent state \cite {TLiu} defined
as $|n\rangle_{\alpha}=(1/\sqrt{n!})(a^{\dagger}+\alpha)^{n}\exp\{-\alpha a^{\dagger}-\alpha^{2}/2\}|0\rangle$.
It is easily checked that $|\Phi^{+}\rangle$ ($|\Phi^{-}\rangle$) is
the eigenfunction with even (odd) parity under the parity operator
$\Pi=\sigma_{x}e^{i\pi a^{\dagger}a}$ with $\Pi
|\Phi^{\pm}\rangle=\pm |\Phi^{\pm}\rangle$. Considering the unitary
transformation $V$ for Eq. (5), we have $\Pi=V^{+}PV$, implying that the
parity operator $P$ commutes with both the Hamiltonian $H_{R}$ under
the RWA and the Hamiltonian $H_{NR}$ beyond the RWA.

Putting Eq. (6) into the Schr\"odinger equation of Eq. (5) yields,
\begin{equation}
\omega(m-\alpha^{2})f_{m} \pm \frac
{\omega_{0}}{2}\sum_{n}f_{n}D_{mn} = Ef_{m},
\end{equation}
where $D_{mn}$ is given by \cite {TLiu}
$$D_{mn}=e^{-2\alpha^{2}}\sum_{k=0}^{\min
[m,n]}(-1)^{-k}\frac{\sqrt{m!n!}(2\alpha)^{m+n-2k}}{(m-k)!(n-k)!k!}.$$
By solving Eq. (7), we may immediately obtain the Berry phases as
\begin{eqnarray}
\gamma^{'\pm}=i\int_{C} d\varphi \langle\Phi^{\pm}|V^{+}U^{+}(\varphi)\frac {d}{d\varphi}
U(\varphi)V|\Phi^{\pm}\rangle  \notag \\
=2\pi \{\alpha^{2} + \sum_{n=0}^{M}[n(f_{n}^{\pm})^{2}-2\alpha\sqrt{n}f_{n}^{\pm}f_{n-1}^{\pm}]\},
\end{eqnarray}
where we have used the prime here to distinguish from the result
under the RWA, and the superscript '+' ('-') is for the even (odd)
parity. $M$ is the integer relevant to the size of the truncated
space spanned by our employed eigenfunctions \cite {TLiu} and
$f_{k}^{\pm}$ $(k=n,n-1)$ is to be determined later.

Despite the possibility to solve Eq. (7) numerically, we would like to solve Eq. (7)
analytically in order to show the
physics more clearly. We consider the expansion with $m,n=k,k+1$, which
is actually the first-order approximation in \cite {TLiu}. This leads to \cite {addition}
\begin{equation}
(E^{\pm})^{2} - (\eta^{\pm}_{k,k}+\eta^{\pm}_{k+1,k+1})E^{\pm} + \eta^{\pm}_{k,k}\eta^{\pm}_{k+1,k+1}
+ \omega_{0}^{2}D^{2}_{k,k+1}/4 = 0,
\end{equation}
with $\eta_{k,k}^{\pm}=\omega(k-\alpha)\mp\omega_{0}D_{k,k}/2$. For $k=0$,
we found that the trivial solution under the RWA is not trivial
any more due to involvement of the counter-rotating terms and it
evolves as the ground state of the system meeting even parity with the
eigenenergy as
$E_{GS}=(1/2)[\eta^{+}_{0,0}+\eta^{+}_{1,1}-\sqrt{(\eta^{+}_{0,0}-\eta^{+}_{1,1})^{2}+
\omega_{0}^{2}D_{0,1}^{2}}]$. Taking
the corresponding eigenfunction, we obtain the VIBP
\begin{eqnarray}
\gamma^{'}_{GS}=2\pi[\alpha^{2}+(f^{+}_{1})^{2}+2\alpha f^{+}_{0}f^{+}_{1}] \notag \\
=2\pi(\alpha^{2}+\frac {1-2\alpha\mu_{0}}{1+\mu_{0}^{2}}), \ \ \ \ \ \ \
\end{eqnarray}
where $f^{+}_{0}=\mu_{0}/\sqrt{1+\mu_{0}^{2}}$ and
$f^{+}_{1}=1/\sqrt{1+\mu_{0}^{2}}$ are coefficients of the ground state wavefunction with
$\mu_{0}=\omega_{0}D_{0,1}/[2(\eta^{+}_{0,0}-E_{GS})]$. The VIBP
$\gamma^{'}_{GS}$ is  never zero or multiple of $2\pi$, much
different from $\gamma_{GS}$ obtained under the RWA.

The original VIBPs are actually the Berry phases relevant to the lowest excited states of
the system. Using an arbitrary integer $k$ (different from zero),
we obtain from Eq. (9) the eigensolutions of the excited states with even or odd parity
(see details in Appendix), and the counterparts of the original VIBPs beyond the RWA
are of odd parity, i.e.,
$\gamma^{'-}_{0,-}=2\pi(\alpha^{2} + \frac {1-2\alpha\mu_{0,-}}{1+\nu_{0,-}^{2}})$ and
$\gamma^{'-}_{0,+}=2\pi(\alpha^{2} + \frac {1-2\alpha\nu_{0,+}}{1+\nu_{0,+}^{2}})$,
with $\mu_{0,-}$ and $\nu_{0,-}$ defined in Appendix.

\begin{figure}[tbph]
\centering
\includegraphics[width=3.0 in]{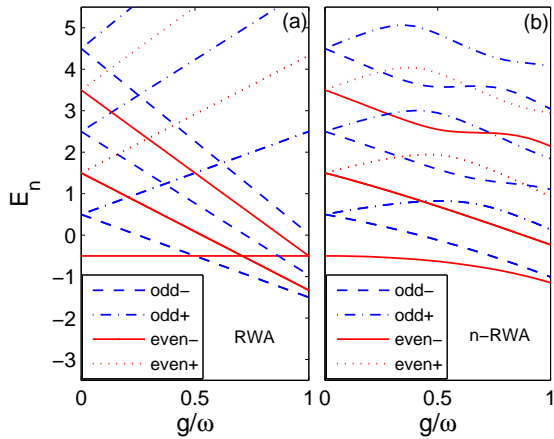}
\par
\caption{(color online) Comparison of the eigenenergies under the RWA (a) with
beyond the RWA (b) depending on the dimensionless coupling $g$, where we
set $\Delta=0$.} \label{fig1}
\end{figure}

The comparison in Fig. 1 for the eigenenergies solved under the RWA
with those beyond the RWA demonstrates the increasing difference with the coupling $g$.
Particularly, it shows that $|\Psi_{GS}\rangle$ under the RWA is really the
ground state of the system with counterpart in the case beyond
the RWA. As a result, it is reasonable for us to define the VIBP based on $|\Psi_{GS}\rangle$.

More importantly, by removing the RWA, we have very different VIBPs
from those under the RWA. As plotted in Fig. 2, our defined VIBP
varies from zero under the RWA to non-zero values beyond the RWA,
and the original VIBPs also change significantly. Although our new
definition of the VIBP might be only of conceptual interest, our
treatment of  the original VIBPs is of practical application.
From Fig. 2 we know that the values of Berry phases strongly depend
on the RWA. As long as $g$ is not strictly zero, the Berry phases
obtained under the RWA are different from those beyond the RWA,
particularly for the case of large detuning $\Delta$. Consequently,
when considering the quantum logic gating based on the Berry phase
in JC-type model \cite {qip1}, employment of the RWA, yielding
inaccurate Berry phase, would introduce errors in the logic gating.
Since the idea in \cite {zanardi}, there have been many proposals
and experiments using Berry phases for universal QIP \cite {qip1,
qip2} because of the robustness of the geometric phases under some
kinds of noise \cite {qip3}. Among those works JC model has been
extensively employed, in which the qubits are encoded in the
spin-1/2 states and the quantized field is used as data bus. For an
efficient operation \cite {qip1}, the light-matter coupling (i.e.,
$g$ used above) is required to be reasonably large, and in some
cases detuning (e.g., $\Delta$ used above) is necessary to be large
enough for removing effectively the quantized field from the qubit
coupling. In terms of our present results, employment of the RWA
should be considered with great caution because the generated
geometric phases would be deviated from the expectation due to
contributions of counter-rotating terms. This is the intrinsic error
in the models involving the RWA, which might be incorrectly
resorted to imperfections due to diabaticity or operations in
experiments.

We have noticed the experimental reports for quantum gating
operations using  geometric phases in trapped ions \cite
{ion-1,ion-2}, in which the RWA was employed in both the JC-type
models. Taking the latter experiment \cite{ion-2} as an example. We
find the Rabi frequency $\Omega=2\pi\times 0.21$ MHz, the Lamb-Dicke
parameter $\eta=0.056$ and the trap frequency being $6\Omega$. This implies
that the measured geometric phase, although under the RWA, is nearly accurate because
the corresponding dimensionless coupling is smaller than 0.01. The
key point in this case is the tiny Lamb-Dicke parameter which
significantly reduces the effective Rabi frequency, although it is
disadvantageous for large-scale QIP. This also means
that the future large-scale trapped-ion QIP using geometric phases, if considering faster operations, should
seriously justify the RWA.

\begin{figure}[tbph]
\centering
\includegraphics[width=3.0 in]{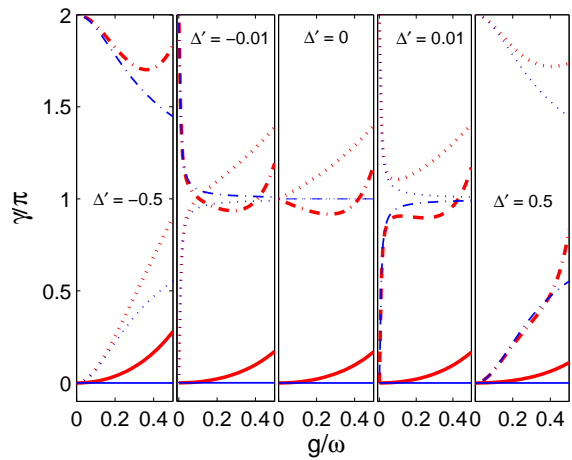}
\par
\caption{(color online) Comparison of the Berry phases under the RWA (blue curves) with
beyond the RWA (red bold curves) vs the dimensionless coupling $g/\omega$ and detuning $\Delta'=\Delta/\omega$, where the solid, dashed
and dashed-dotted curves are for $\gamma_{GS} (\gamma'_{GS})$, $\gamma_{0-}^{-} (\gamma_{0-}^{'-})$, and $\gamma_{0+}^{-} (\gamma_{0+}^{'-})$, respectively. } \label{fig2}
\end{figure}

Our result about the VIBPs seems opposite to that in a recent paper
\cite {larson} which pointed out that the original VIBPs beyond
the RWA are strictly zero. The main difference of our work from
\cite {larson} is the fully quantum mechanical treatment presenting
reasonable comparison between the solutions with and without the
RWA, which enables us to understand the contribution from the
counter-rotating terms. We have demonstrated in Figs. 1 and 2 the
gradually increasing differences with the coupling $g$ due to
involvement of the counter-rotating terms. The contribution from the
counter-rotating terms is also reflected in our defined VIBP, which,
beyond the RWA, is associated with
$|n\rangle_{\alpha}$ and $|n\rangle_{-\alpha}$, instead of the
vacuum state of the quantized field any more. This is due to
additional excitations of the system by the counter-rotating terms,
which enhances the Berry phases and also makes the situation more
complicated. In contrast, in the semi-classical treatment in \cite
{larson}, the addition of the counter-rotating terms cancels the
original Berry phases, for which it is not clear whether the changes
of the VIBPs are sudden or continuous with deletion of the RWA. In addition, it is also not clear in the
semi-classical treatment to which quantized field states the
considered Berry phases correspond.

In summary, we have defined a new VIBP based on the eigensolutions
of JC model and  investigated analytically the VIBP and the
original VIBPs by removing the RWA from the JC
model. From the analytical expressions, we might be able to
understand how and what physical parameters bring about the changes,
and it also helps us to realize that justification of employing the
RWA should be seriously taken by considering both energies of the
system and the geometric property of the state evolution.
Particularly, more attention should be paid to identification of the
geometric phase if we employ a JC model for designing quantum gating
schemes. Since the JC model has been widely employed in different fields and
is essential to QIP designs, our results should be of general interests in both concept
and application. Moreover, our study on deletion of the RWA is not
restricted to the condition of adiabaticity. So we believe that our
results could be applied to both adiabatic geometric phase - Berry
phase, and non-adiabatic geometric phase, such as the
Ahaonov-Anandon phase \cite {AA}. In one word, our results open up a
new arena to the study of breakdown of the RWA in the geometric
evolution of quantum systems.

This work is supported by NNSFC under Grant No. 10974225 and by funding from WIPM.

\section*{Appendix}
We compare the eigensolutions of the JC model in the presence and absence of the RWA.

\textbf{Under the RWA} ~
For Eq. (1), the eigenfunctions conventionally considered are
$|\Psi_{n,\pm}\rangle$ in Eq. (3) with  corresponding eigenenergies
$E_{n,\pm}=(n+1/2)\omega\pm\sqrt{\Delta^{2}+4g^{2}(n+1)}/2$. In
fact, there is another eigenstate decoupled from the interaction,
i.e., $|\Psi_{GS}\rangle=|0\rangle|g\rangle$ with corresponding
eigenenergy $E_{G}=-\omega_{0}/2$.

Introducing the parity operator $P=-\sigma_{z} e^{i\pi
a^{\dagger}a}$, we  have $[P, H_{R}]=0$ with
$P|\Psi_{n,\pm}\rangle=(-1)^{n+1}|\Psi_{n,\pm}\rangle$ and
$P|\Psi_{GS}\rangle=|\Psi_{GS}\rangle$. So the
eigenfunctions are not only related to the upper or lower level of the spin-1/2, but
also associated with the even or odd parity.

\textbf{Beyond the RWA} ~
The first-order expansion of Eq. (7) leads to Eq. (9), from which
we may obtain  analytical eigensolutions of Eq. (5). Except the
ground state written in the text, the excited states, like in
the RWA situation, are expressed to be related both to the upper or lower
level, and to the even or odd parity. For the excited eigenstates
with even parity, the eigenenergies are
\begin{eqnarray}
E^{+}_{k,\pm}=(1/2)[\eta^{+}_{k,k}+\eta^{+}_{k+1,k+1}\pm \notag \\
\sqrt{(\eta^{+}_{k,k}-\eta^{+}_{k+1,k+1})^{2}+
\omega_{0}^{2}D_{k,k+1}^{2}}],   \label{A1}
\end{eqnarray}
corresponding to eigenfunctions with coefficients
$f^{+}_{k,\pm}=\mu_{k,\pm}/\sqrt{1+\mu_{k,\pm}^{2}}$,
$f^{+}_{k+1,\pm}=1/\sqrt{1+\mu_{k,\pm}^{2}}$, and
$\mu_{k,\pm}=\omega_{0}D_{k,k+1}/[2(\eta^{+}_{k,k}-E_{k,\pm}^{+})]$.
The corresponding Berry phases are $\gamma^{'+}_{k,\pm}=2\pi(\alpha^{2} + k - \frac {1-2\alpha\sqrt{k+1}\mu_{k,\pm}}{1+\mu_{k,\pm}^{2}}).$

In contrast, for the excited eigenstates with odd parity, the eigenenergies are
\begin{eqnarray}
E^{-}_{k,\pm}=(1/2)[\eta^{-}_{k,k}+\eta^{-}_{k+1,k+1}\pm\notag \\ \sqrt{(\eta^{-}_{k,k}-\eta^{-}_{k+1,k+1})^{2}+
\omega_{0}^{2}D_{k,k+1}^{2}}],  \label{A3}
\end{eqnarray}
corresponding to eigenfunctions with coefficients
$f^{-}_{k,\pm}=\nu_{k,\pm}/\sqrt{1+\nu_{k,\pm}^{2}}$,
$f^{-}_{k+1,\pm}=1/\sqrt{1+\nu_{k,\pm}^{2}}$, and
$\nu_{k,\pm}=-\omega_{0}D_{k,k+1}/[2(\eta^{-}_{k,k}-E_{k,\pm}^{-})]$.
The corresponding Berry phases are $\gamma^{'-}_{k,\pm}=2\pi(\alpha^{2} + k + \frac {1-2\alpha\sqrt{k+1}\nu_{k,\pm}}{1+\nu_{k,\pm}^{2}}).$


\begin{thebibliography} {99}
\bibitem{addition} The first-order approximation yields a two-by-two matrix relevant to Eq. (7). It works well for arbitrary values of $g$ \cite {TLiu} and
is more concise than higher-order approximation to show physics beyond the RWA.
\end{thebibliography}
\end{document}